\documentstyle[a4wide,epsf,11pt,titlepage]{article}

\newcounter{nref}
\newcommand{\bbib}{%
  \renewcommand{\refname}{\large\bf References}%
  \begin{thebibliography}{99}%
  \setcounter{enumiv}{\thenref}}
\newcommand{\ebib}{%
  \end{thebibliography}%
  \setcounter{nref}{\arabic{enumiv}}}
\newcommand{\head}[3]{%
  \setcounter{nref}{0}%
  \thispagestyle{empty}%
  \section*{\LARGE\bf #1}%
  \stepcounter{section}%
  \addcontentsline{toc}{section}{#1}%
  \large\itshape%
  #2\\\vspace{0.1pt}\\%
  #3%
  \normalsize\upshape%
  \bigskip}

\begin{document}

%----------------------------------------------------------------------

\head{Understanding Core-Collapse Supernovae}
     {A. Burrows$^1$}
     {$^1$Steward Observatory and the Department of Astronomy, University of Arizona, Tucson, AZ USA 85721}

At the University of Arizona, we have a major effort to study core-collapse
supernovae, neutrino opacities and transfer, multi-dimensional radiation hydrodynamics, pulsar kicks,
and gravitational radiation from supernovae.  We have developed a precision
1D (spherical) multi-group implicit Boltzmann solver, SESAME
\footnote{{\bf S}pherical, {\bf E}xplicit/Implicit, {\bf S}upernova,
{\bf A}lgorithm for, {\bf M}ulti-Group/Multi-Angle,
{\bf E}xplosion Simulations} \cite{aburrows.young,aburrows.sesame},
using the Feautrier technique, with the tangent-ray method to achieve excellent angular
resolution.  This code provides the spherical benchmark and testbed for our multi-D
efforts (e.g., VULCAN/2D, see below). 

In an effort to explore non-standard effects quickly and heuristically, 
Thompson, Quataert, and Burrows\cite{aburrows.tqb} (TQB) have modified SESAME to incorporate
the centrifugal effects of differential rotation in an approximate way, as well as the heating and angular
momentum redistribution effects of viscosity (perhaps due to the magneto-rotational
instability, MRI).  The neutrino-driven mechanism is already close
to explosion and requires an increase in the energy deposition rate in the ``gain
region" of only about 25\% to 40\%.  TQB have
shown that rapid rotation of the initial Chandrasekhar core can provide
such a boost through the agencies of centrifugal support and viscous heating alone.  
As the explosion commences, residual neutrino cooling subsides and neutrino heating
takes over, but the initial boost due to viscous heating in these calculations helped trigger explosion.
They have also shown that a delayed explosion more naturally ejects high Y$_e$ material
($\sim 0.5$) than a prompt (never achieved in realistic simulations) or early
delayed explosion.  Furthermore, the larger residual core mass from which the neutron star is formed
is more in keeping with larger measured pulsar masses.  However, a core that explodes
using a viscous boost is also rotating more rapidly than
the inferred birth spin rates of known pulsars.  Hence, a mechanism that sheds angular
momentum during and after explosion is required.  TQB show that here too the viscosity
that heats the core can also lead to the diffusion of angular momentum out of the core
(by a factor of $\sim$5 by the end of their calculations) on a timescale of seconds.
These calculations, being spherical, are by no means definitive, and true multi-D
simulations, including magnetic fields, are required to answer the questions the
work of TQB pose.  However, these results are intriguing.

As a biproduct, TQB have also estimated the evolution of a core's rotational profile during collapse
and after bounce, i.e. the mapping between the initial and the ``final" angular
velocities.  This study with neutrino transport complements that of Ott et al.\cite{aburrows.ott}
without neutrino transport (which nevertheless explored a wide range of initial conditions).

As TQB show, a useful criterion for explosion is that the timescale ($\tau_{adv}$)
for matter to flow from the stalled shock into the cooling region near the neutrinosphere
is larger than the characteristic heating timescale ($\tau_{H}$).  This criterion
is loosely equivalent to the $L_{\nu}$ vs. $\dot{M}$ criterion 
of Burrows \& Goshy\cite{aburrows.goshy}. For a given $\dot{M}$, there is a threshold $L_{\nu}$,
above which explosion ensues.  That threshold is lower in 2D/3D than in 1D.  In 1D, $\tau_{adv}$
is too small and matter settles onto the core before sufficient heat is deposited
by the neutrinos.  However, as shown by Burrows, Hayes, \& Fryxell\cite{aburrows.bhf} (BHF),
Herant et al.\cite{aburrows.herant}, and Janka  and M\"uller\cite{aburrows.janka}, in 2D with the neutrino-driven
overturning instability and convective motions behind the
stalled shock enabled, $\tau_{adv}$ can be larger than $\tau_{H}$.  The swirling motions
keep the matter in the gain region longer, increasing $\tau_{adv}$.  As a
result, the 2D calculations of BHF and Herant et al., without rotation,
led to explosions.  Note that if matter did not settle onto the core at all,
or if the cooling rates near the neutrinospheres interior to the gain region
were low, explosion by neutrino heating would naturally, and trivially, result.
We know how to obtain explosions.  The question is whether Nature agrees.

Importantly, the results of BHF and Herant et al. were obtained using simple
neutrino transfer and transport algorithms.  Though they may have been
substantially correct, it is thought that better calculations
of the neutrino-matter coupling, particularly in the semi-transparent
gain region, are required to falsify and/or check the convective neutrino-driven
mechanism of explosion that I, for one, think obtains.  Hence, a number of groups (for instance, at
Arizona, MPA, and ORNL) are developing more sophisticated multi-dimensional, multi-group
neutrino transfer schemes and in the next few years we should be able
to make a much more detailed assessment of this mechanism.  

\begin{figure}[ht]
   \centerline{\epsfxsize=0.8\textwidth\epsffile{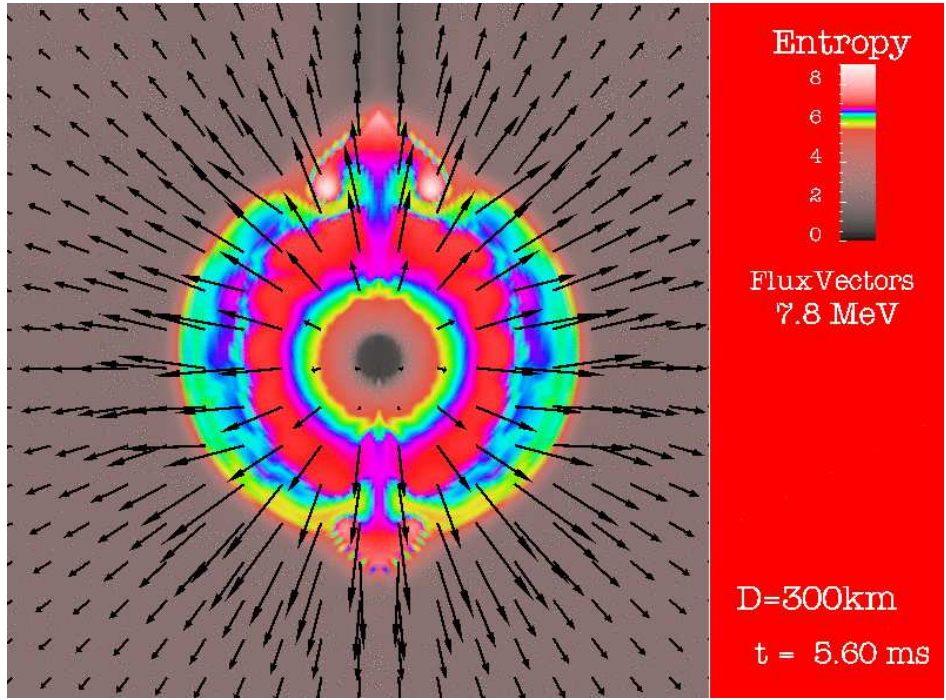}}
  \caption{A snapshot 5.6 milliseconds after bounce.  The 2D MGFLD variant
of VULCAN/2D was used.  The color map is of the entropy and
the vectors are the electron-neutrino differential fluxes at
7.8 MeV. The core is not rotating, but in
the beginning of the calculation it was given a modest
anisotropic perturbation in the density at the
north pole. The approximate scale is 300 kilometers.
(Obtained in collaboration with R. Walder.)}
  \label{aburrows.fig1}
\end{figure}

In this spirit, we \cite{aburrows.livne} have developed the code ``VULCAN/2D"
and are testing it in full Boltzmann and MGFLD modes.  Figure \ref{aburrows.fig1}
provides a snapshot in a 2D MGFLD run of the electron 
neutrino flux vector field at 7.8 MeV, superposed
on an entropy color map.  VULCAN/2D is an
ALE code with remapping that implicitly solves the 6-dimensional problem (1(time) + 2(space) +
2(angles) + 1(energy-group)) using the S$_n$ method to tile
angular space.  It is parallelized using MPI, is flux-conserving, and smoothly matches to the diffusion regime,
but does not include the Doppler and aberration velocity terms.  The former
is important, particularly around bounce and will be included explicitly
in operator-split fashion in a later version. However, the achievement of the current VULCAN/2D is
a milestone, being the first such time-dependent code available
in core-collapse studies with truly ``2D", as  opposed
to ray-by-ray or only spherical, capability.  Ray-by-ray
codes, such as used by BHF, Rampp \& Janka\cite{aburrows.rampp}, 
and Buras et al.\cite{aburrows.buras}, can exaggerate the variation
in the neutrino energy density and flux with angle by as much as 20\%-50\%,
with the concommitent artificial variation in the heating rates with angle.
However, it is not yet known how, in the integral sense 
of global heat deposition, this difference might effect
the mechanism itself.

\begin{figure}[ht]
   \centerline{\epsfxsize=0.8\textwidth\epsffile{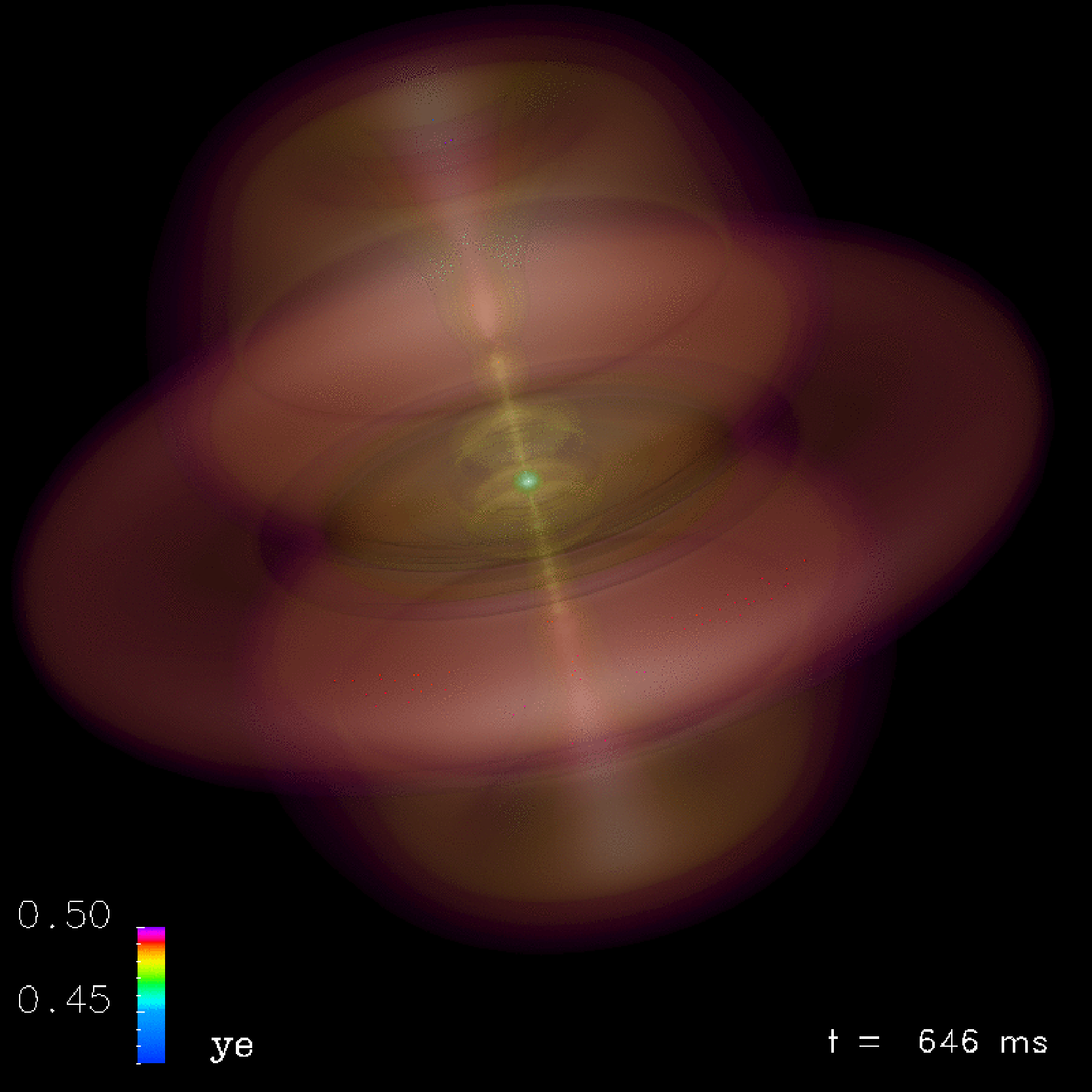}}
  \caption{A snapshot after bounce of a calculation of a 20 M$_{\odot}$ model with rotation.
The surfaces are nested isodensity shells and the approximate scale is 600 kilometers.
Funnels due to an emerging centrifugal barrier are clearly seen along the poles,
as is an equatorial bulge. (Obtained in collaboration with C. Ott and R. Walder.)}
  \label{aburrows.fig2}
\end{figure}

Whether rotation is central to the mechanism of core-collapse supernovae
is not known.  As the work of BHF, Herant et 
al.\cite{aburrows.herant}, and Janka and M\"uller\cite{aburrows.janka}
suggest, it may not be.  However, as TQB, Ott et al.\cite{aburrows.ott}, 
and Fryer \& Heger\cite{aburrows.fryer} demonstrate, rotation (naturally
amplified during collapse and compaction) introduces interesting new phenomena.
In addition to decreasing the effective gravity against which the explosion
must emerge, enlarging the gain region, and generating funnels along the poles
through which an explosion can more easily pierce\cite{aburrows.bom}, 
stars do in fact rotate.  Unfortunately, how much the Chandrasekhar core rotates just before
collapse is currently a mystery.  An example of such post-bounce funnel structures
is given in Fig. \ref{aburrows.fig2}. (I refer the reader to the bi-polar
structure that HST has revealed in SN1987A.)  Therefore, the consequences
of the fact of rotation, and the possible synergistic effects with 
magnetic fields\cite{aburrows.tqb,aburrows.akiyama}, 
deserve further scrutiny in the context of the neutrino-driven
paradigm of core-collapse supernova explosions.   

\subsection*{Acknowledgements}

I am happy to acknowledge collaboration with Rolf Walder,
Christian Ott, Eli Livne, Todd Thompson, Itamar Lichtenstadt,
and Jeremiah Murphy on various of the topics mentioned in 
this abstract. Support for this work is provided in part by
the SciDAC program
of the US DOE, grant number DE-FC02-01ER41184.  

\def\apj{{\em Ap.~J.}\ }
\def\aap{{\em Astr.~Ap.}\ }

%----------------------------------------------------------------------

\end{document}